# The Successful Operation of Hole-type Gaseous Detectors at Cryogenic Temperatures

L. Pereiale[1,2], V. Peskov[3], C. Iacobaeus[4], T. Francke[5], B. Lund-Jensen[6], P. Pavlopoulos[1,3], P. Picchi[1,2], F. Pietropaolo[1,7], F.Tokanai[8]

[1] CERN, Geneva, Switzerland
[2] IFSI-CNR of Torino, Torino, Italy
[3] Leonardo de Vinci University, Paris, France
[4] Karolinska Institute, Stockholm, Sweden
[5] XCounter AB, Danderyd, Sweden
[6] Royal Institute of Technology (KTH), Stockholm, Sweden
[7] INFN Padova, Italy
[8] Yamagata University, Yamagata, Japan

*Abstract--*We have demonstrated that hole-type gaseous detectors, GEMs and capillary plates, can operate up to 77 K. For example, a single capillary plate can operate at gains of above $10^3$ in the entire temperature interval between 300 until 77 K. The same capillary plate combined with CsI photocathodes could operate perfectly well at gains (depending on gas mixtures) of 100-1000. Obtained results may open new fields of applications for capillary plates as detectors of UV light and charge particles at cryogenic temperatures: noble liquid TPCs, WIMP detectors or LXe scintillating calorimeters and cryogenic PETs.

## I. Introduction

In several studies and applications there are needs for the detection of VUV light at cryogenic temperatures. Examples could be some high energy physics and astrophysics experiments, noble liquid PETs, studies of cryogenic plasmas and studies of quantum phenomenas in liquid and solid He. If we focus only on high energy physics then one of the main applications could be noble liquid TPCs and noble liquid scintillation calorimeters.

TPCs are rather widely used detectors: the ICARUS experiment, the nTOF experiment and WIMP search LXe/Ar detectors are all examples. A cryogenic TPC is usually a large volume filled with noble liquids (see Fig.1). If an interaction occurs inside the liquid scintillation light and a charge track are produced simultaneously. The scintillation light is detected by photodetectors. By influence of the electric field, parts of the primary electrons from the track are collected on the position sensitive readout plate. This allows one to obtain a 2D projection of the track. The time difference between the scintillation light and charge collection on the position sensitive elements is used for the 3D track reconstruction. A noble liquid TPC is a very powerful detector allowing one not only to visualize the tracks, but also to measure the deposited energy and perform event rejection from the measured light to the charge ratio. As one can see photodetectors are important components of the TPCs. Usually PMs are used as photodetectors.

In a recent publication [1] we presented some preliminary results demonstrating that gaseous detectors with solid photocathodes can operate at low temperatures. This may open possibilities of replacing costly and bulky PMs with cheap and simple photosensitive gaseous detectors. Note also that gaseous detectors are not sensitive to magnetic fields desirable for some experiments.

In this paper we will present our new results of systematic studies of the operation of gaseous detectors at cryogenic temperatures obtained with the latest equipments.

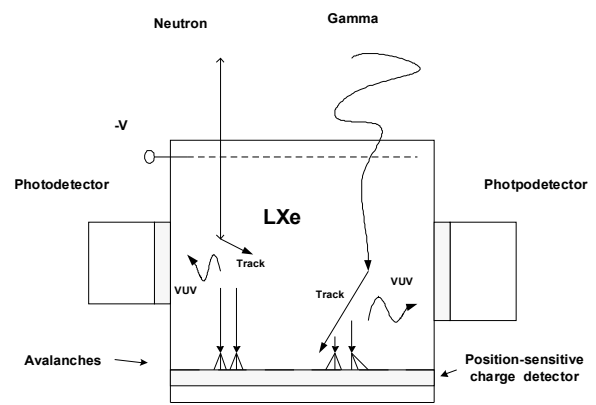

Fig. 1 A schematic drawing illustrating the operational principle of a noble liquids TPC.

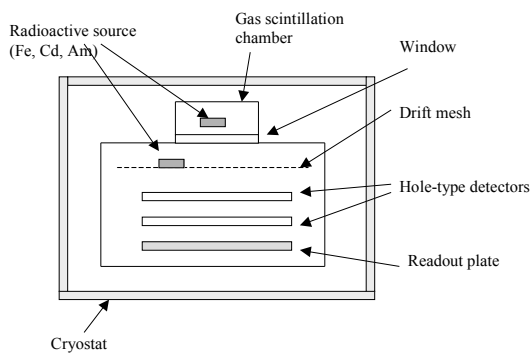

Fig. 2. A schematic drawing of the experimental set-up for measurements with radioactive sources.

## II. Experimental Set Up

Our experimental chamber is shown schematically in Fig. 2. It consists of a cryostat with a test chamber located inside (see also Fig.3). The cryostat we used was originally designed for tests of ATLAS calorimeter modules. It was a very helpful device in these studies since it was computer controlled and allowed the cooling to take place to $LN_2$ or any other temperature above it with a selected speed profile. The test chamber was filled with $Ar+10\%CH_4$ or $He+10\%H_2$ gas mixture at a pressure of 1 atm. If necessary a gas scintillation chamber could be attached to the test chamber via a $MgF_2$ window (see Fig.2). It was filled with Ar or Kr at P=1 atm. The scintillation light was produced by an alpha source placed inside. Inside the test chamber various gaseous detectors could be installed. The main focus in these studies was on the hole-type detectors because they have unique properties which are: the ability to operate in poorly quenched gas mixtures and the possibility to work in cascade mode. The hole-type detectors used in these studies were GEMs or capillary plate (CPs). GEMs had sizes of 5x5 cm² and hole pitches (50 µm in diameter) of 140 µm. GEMs were used either in CAT (the cathode was electrically connected to the readout plate placed in contact) or cascade (two GEMs) mode. All capillaries used had diameters of 20 mm, thicknesses of 0.4 or 0.8 mm, and diameters of capillary holes of 40 or 100 µm.

We also studied the operation of these detectors in combination with CsI and others solid photocathodes (see [2] for more details). For these measurements the test chamber was slightly modified as shown in Fig.4. In this paper we will present only some selected results obtained with reflective and semitransparent CsI photocathodes. Semitransparent CsI (20 nm thick) was deposit on the inner surface of the $MgF_2$ window. In the case of the reflective CsI photocathode, it was deposited on the top electrode of the hole–type detector (the thickness was 400 nm).

For feedback measurements a pulsed $H_2$ lamp (a few ns) was placed outside the cryostat (see Fig. 4). The same lamp in combination with a Hg lamp was also used for the quantum efficiency measurements.

The signals from the capillaries were read out by a charge sensitive amplifier Ortec, 142 PC. In some measurements we also recorded pulse-height spectra produce by radioactive source or by $H_2$ lamp (see [2] for more details).

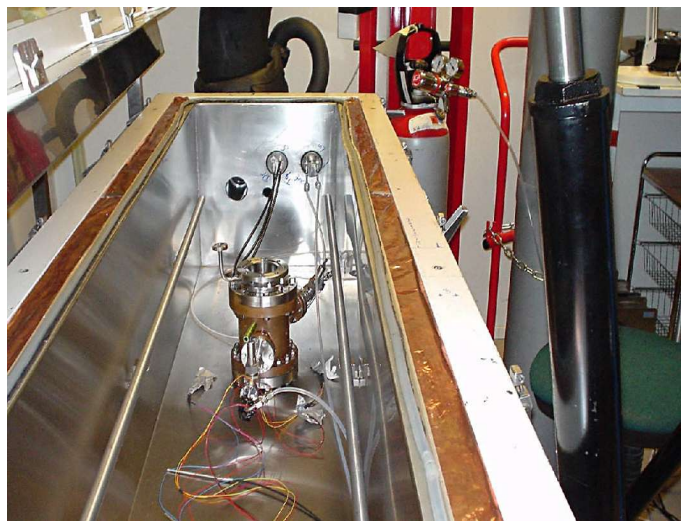

Fig.3. A photo of the cryostat with the test chamber inside

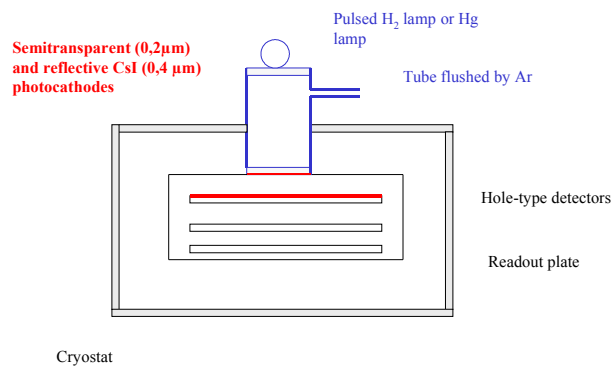

Fig. 4. Set-up modifications (in bleu and red) for measurements with external UV sources.

## III. Results

*III. 1 Gain Measurements with Radioactive Sources*

Figs. 5 and 6 show the gain measurements of a single GEM operating in $Ar+CH_4$ and $He+H_2$ gas mixtures at P=1atm. One can see that the maximum achievable gain of the GEM drops sharply with the temperature, however, it could still operate even at an LN$_2$ temperature.

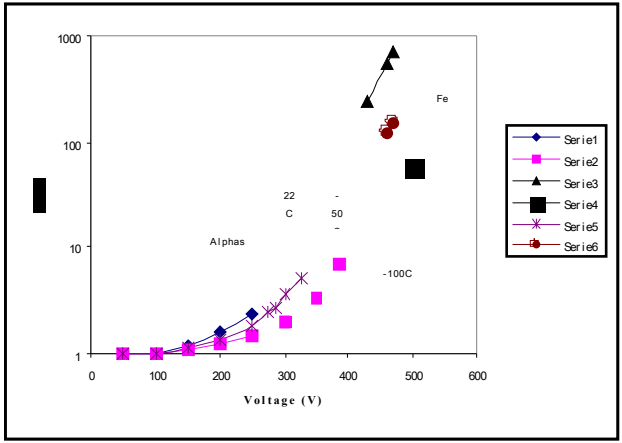

Fig. 5. GEM gains (CAT mode) measured with alphas and Fe in Ar+10%CH$_4$ at various temperatures.

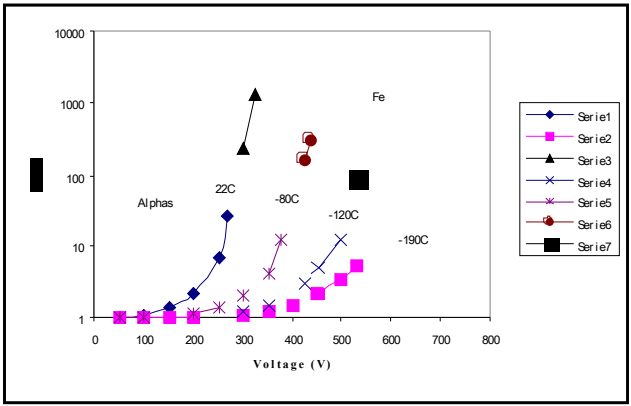

Fig.6. GEM gains (CAT mode) measured with alphas and Fe in He+10%H$_2$ at various temperatures.

The figures 7 and 8 show the gain of the CPs in He+H$_2$ and Ar+CH$_4$ mixtures and in pure Ar at various temperatures. One can see that the maximum achievable gain of the CPs is much higher than for the GEMs, although it drops with the temperature as well. Note that CPs could operat at gains of 10$^3$ even in pure Ar (see [2] for more details).

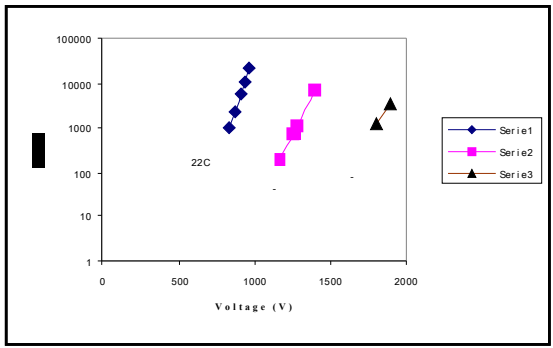

Fig. 7 CP gains (Fe) in He+H$_2$.

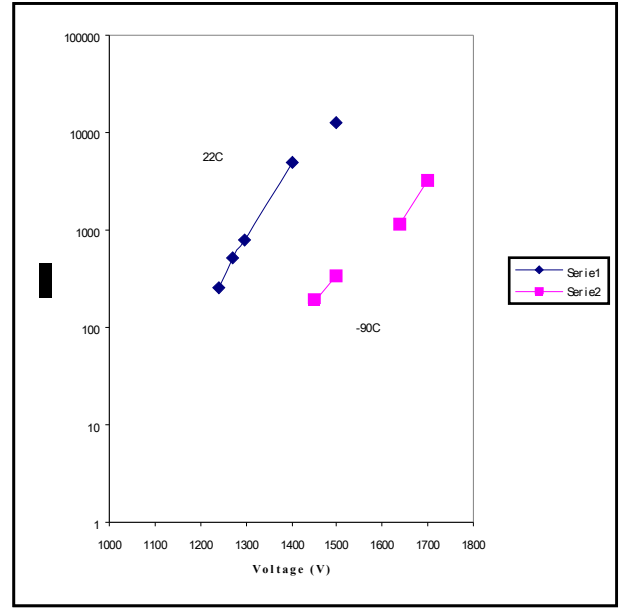

Fig 8. Gains (Fe) of conventional CPs in Ar+CH$_4$.

We also tested the operation of H$_2$ treated CPs. Advantage of this type CP is ability to operate at high counting rate [3]. Figure 9. shows how the resistivity changes with the temperature. The resistivity measured between the electrodes of the H$_2$ treated CP was ~1GΩ. At this temperature the resitivity obeyed the Ohm's law. At LN$_2$ temperature, the resistivity reaches the value of ~100 GΩ and begins to deviate from Ohm's law. Gain curves for the H$_2$-treated CPs are presented in Fig. 10.

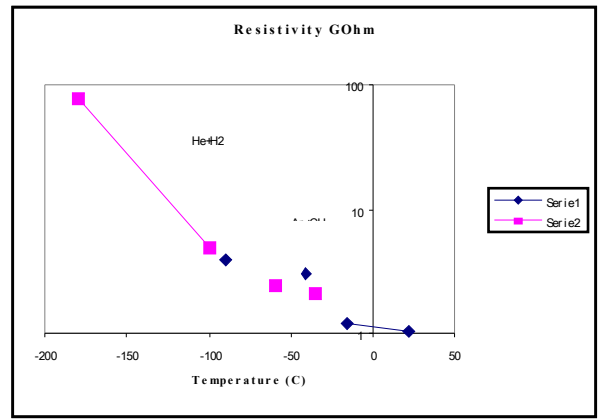

Fig. 9. Resistivity of H$_2$- treated CP vs. temperature in Ar+CH$_4$ and He+H$_2$ gas mixtures.

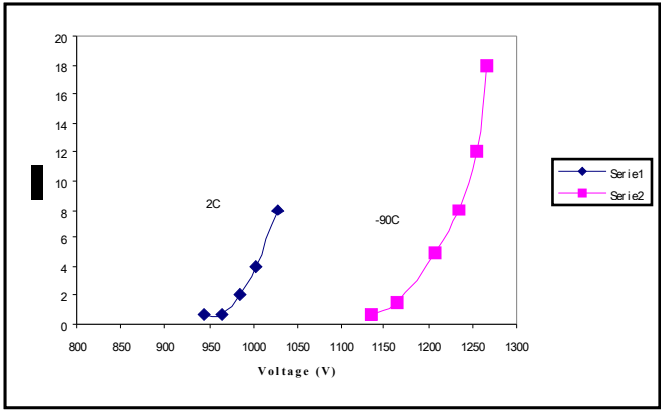

Fig .10.Alpha signals measured on the readout plate
(in charge extraction mode).

*III. 2. The quantum efficiency and Gain Measurements with Hole-type Gaseous Detectors Combined with Reflective and Semitransparent Photocathodes.*

The procedure of the CsI quantum efficiency (QE) calibration is described in [1].

The QE of hole-type detectors in Ar+10%CH$_4$ and He+10%H$_2$ gas mixtures and at various temperatures are presented in Fig. 11. One can see that the value of the QE was, depending on the detector, between 14 and 5 %. Note that: the loss of the measured efficiency at T~80 K is due to the strong back diffusion of photo electrons in the He+H$_2$ gas mixture [1]. The absolute QE of the CsI itself remains almost unchanged with temperature [1].

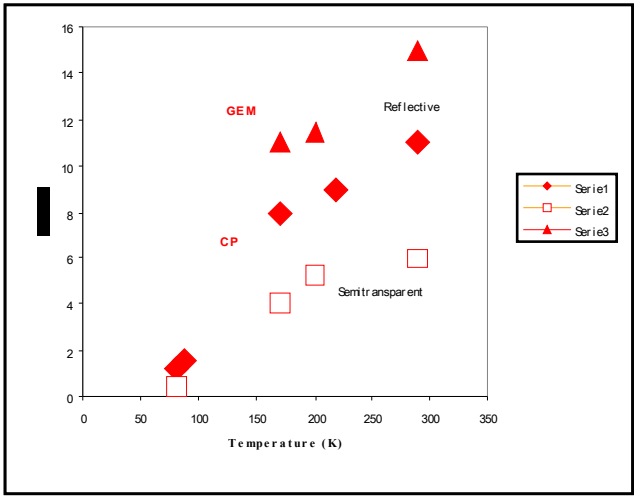

Fig. 11. The QE (Y-axis) of hole- type detectors measured in Ar+10%CH$_4$ and He+10%H$_2$ gas mixtures at various temperatures: triangles-GEM with reflective CsI photocathode, rhombus- CPs with reflective CsI photocathode, open squares- CPs with semitransparent CsI photocathode.

Figs. 12 and 13 summarizes our results obtained with GEMs and CPs. One can see that in both cases the maximum achievable gain of detectors drops with the temperature, but the maximum achievable gain with the CsI photocathode remains almost unchanged. Thus photosensitive gaseous detectors are able to operate at high gains even at cryogenic temperatures.

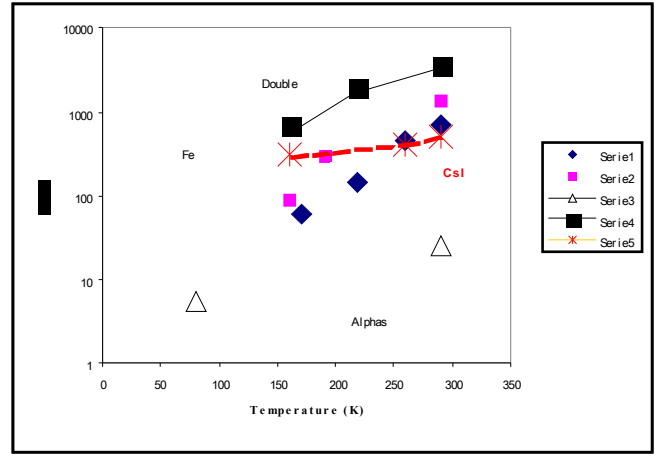

Fig. 12. A general plot: GEM gains (Y-axis) with and without CsI photocathodes. Black squares- double GEM in Ar+CH$_4$ (Fe); rose squares- single GEM in He+H$_2$ (Fe); blue rhombus- single GEM in Ar+CH$_4$ (Fe); open triangles- single GEM in He+H$_2$ (alpha); red stars- double GEM with reflective CsI photocathode.

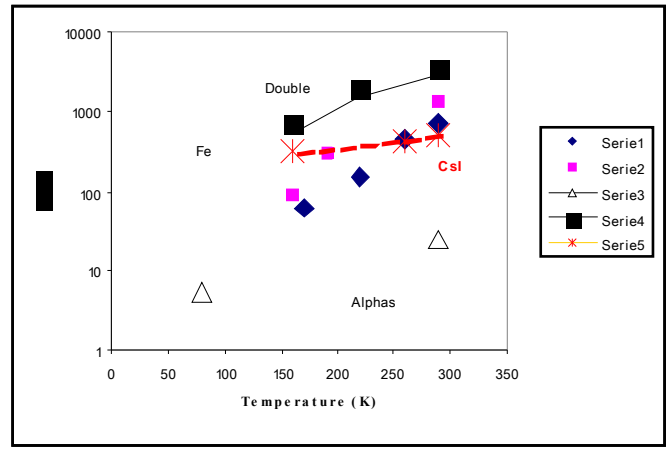

Fig. 13. General plot: CP's gains (Y-axis) with and without CsI photocathodes. Rose squares- CPs in He+H2 (Fe); bleu rhombus- CPs in Ar+CH$_4$ (Fe); open squares-CPs in Ar (Fe); red stars-CP with a semitransparent CsI; red triangles- CPs with reflective CsI.

**IV. Conclusions.**

It was demonstrated experimentally that both GEMs and CPs could operate at cryogenic temperatures-until 78K. CPs offer the highest gain possible in a single step operating mode, thus they could be used to detect charges at cryogenic temperatures, for example in WIMP detectors operating in a charge extraction mode (see also [4]).

The other important discovery was that GEMs and CPs combined with the CsI photocathodes operate at cryogenic temperatures as well, however the maximum achievable gain is lower.

The value of the QE at cryogenic temperatures was reasonably high in order to use these detectors as alternative to PMs. This may open new avenues in applications: such as the readout of noble liquid TPCs.

Note that photosensitive gaseous detectors were already successfully used in the detection of the scintillation lights from noble liquids and gases (see [1] ).